\documentstyle[12pt,psfig,axodraw,a4]{article} 
\def\bild#1#2{    
        \vspace*{-5mm}
        \begin{center}
        \begin{math}
        \epsfxsize#2cm
        \epsffile{#1}
        \end{math}
        \end{center}
        }
\newcommand{\vs}{\vspace{-0.25cm}}

\begin{document} 

\hfill TUM/T39-01-12

\begin{center}
\large{\bf CHIRAL DYNAMICS AND NUCLEAR MATTER}\footnote{Work supported in part 
by BMBF, GSI and DFG.} 

\bigskip 

\bigskip

N. Kaiser$\,^a$, S. Fritsch$\,^a$ and W. Weise$\,^{a,b}$\\

\bigskip

{\small
$^a$\,Physik Department, Technische Universit\"{a}t M\"{u}nchen, D-85747
Garching, Germany\\

\smallskip

$^b$\, ECT$^*$, I-38050 Villazzano (Trento), Italy\\

\smallskip

{\it email: nkaiser@physik.tu-muenchen.de}}

\end{center}

\bigskip

\begin{abstract}
We calculate the equation of state of isospin-symmetric nuclear matter in the 
three-loop approximation of chiral perturbation theory. The contributions to 
the energy per particle $\bar E(k_f)$ from one- and two-pion exchange diagrams 
are ordered in powers of the Fermi momentum $k_f$ (modulo functions of $k_f
/m_\pi$). It is demonstrated that, already at order ${\cal O}(k_f^4)$, two-pion
exchange produces realistic nuclear binding. The underlying saturation 
mechanism is surprisingly simple (in the chiral limit), namely the combination
of an attractive $k_f^3$-term and a repulsive $k_f^4$-term. The empirical 
saturation point and the nuclear compressibility $K\simeq 250\,$MeV are well 
reproduced at order ${\cal O}(k_f^5)$ with a momentum cut-off of $\Lambda
\simeq 0.65$\,GeV which parametrizes short-range dynamics. No further 
short-distance terms are required in our calculation of nuclear matter. In the 
same framework we calculate the density-dependent asymmetry energy and find 
$A_0\simeq 34\,$MeV at the saturation point, in good agreement with the 
empirical value. The pure neutron matter equation of state is also in fair 
qualitative agreement with sophisticated many-body calculations and a 
resummation result of effective field theory, but only for low neutron 
densities $\rho_n <0.25\,$fm$^{-3}$.  
\end{abstract}


\vskip 1.5cm

\section{Introduction}

One of the basic problems in nuclear physics has traditionally been to develop 
a microscopic understanding of the nuclear matter equation of state and the 
properties of finite nuclei in terms of the underlying free nucleon-nucleon 
interaction. The present status is that a quantitatively successful description
of nuclear matter can be achieved, using advanced many-body techniques, in a 
non-relativistic framework when invoking an adjustable three-body force 
\cite{nonrel}. Alternative relativistic approaches treat nucleons as 
Dirac-particles. The simplest one in this category is the $\sigma\omega$-mean 
field model of Serot and Walecka \cite{walecka}. It describes nucleons as
Dirac-quasiparticles moving in self-consistently generated scalar and vector 
mean fields. With two couplings adjusted to the empirical saturation point, 
the nuclear compressibility comes out however much too large (by a factor of 
2). Refinements of relativistic mean field models which include additional 
non-linear terms with adjustable parameters, are nowadays widely used for the 
calculation of nuclear matter properties and finite nuclei \cite{ring}. The 
more basic Dirac-Br\"uckner approach of Brockmann and Machleidt \cite{rolf} 
solves a relativistically improved Bethe-Goldstone equation with a realistic 
nucleon-nucleon interaction parametrized in terms of one-boson exchange 
potentials.

During the last decade, a novel approach to the NN-interaction based on
effective field theory (in particular chiral perturbation theory) has emerged
\cite{ksw,epel,nnpap1,nnpap2}. The key element is a power counting 
scheme which relies on the separation of long and short distance dynamics.     
The methods of effective field theory have recently been applied to systems at
finite density in refs.\cite{furn,hammer}. These works deal mainly with the
effective range expansion (and generalizations thereof) of the equation of 
state of a dilute many-fermion system. The complete resummation of in-medium
multi-loop diagrams for a system with an unnaturally large scattering length 
(such as neutron matter) has been achieved in ref.\cite{steele} (in the limit 
of infinite space-time dimensions). Lutz et al. \cite{lutz} have calculated the
equation of state of nuclear matter from chiral pion-nucleon dynamics. The 
contributions of one- and two-pion exchange diagrams to the energy per particle
have been evaluated up to fourth order in small momenta. An essential
ingredient in their work is an additionally introduced attractive zero-range 
NN-contact interaction. The saturation point obtained in this calculation 
depends very sensitively on the numerical value of $g_A^2/2+g_0+g_1$, the 
strength of this NN-contact interaction. The dynamical origin of the important
attractive NN-contact interaction  remains unexplained. Furthermore, the two 
components of the contact interaction, the one proportional to $g_A^2/2$ and 
the other one parametrized by a coupling $g_0+g_1$, have not been treated on 
equal footing in ref.\cite{lutz} when going to higher orders in perturbation
theory.  

In its starting point our approach to the nuclear matter problem is closely 
related to the work of ref.\cite{lutz}. We also use the three-loop 
approximation of chiral  perturbation theory at finite density to calculate the
contributions of one- and two-pion exchange diagrams to the energy per 
particle of isospin symmetric nuclear matter. In a systematic non-relativistic
expansion we compute all terms up-to-and-including fifth order in small momenta
(these are the Fermi momentum $k_f$ and the pion mass $m_\pi$) and thus go one 
order beyond the work of ref.\cite{lutz}. A major difference in comparison with
ref.\cite{lutz} is that we do not introduce any zero-range NN-contact
interactions with adjustable strengths in addition to the chiral pion-exchange
terms. Instead we use a momentum-space cut-off $\Lambda$ to  regularize the few
divergent parts associated with the chiral two-pion exchange, and we keep the
power divergences which are specific for cut-off regularization. Their 
contribution to the energy per particle is equivalent to that of a zero-range 
contact interaction parametrized through the scale $\Lambda$, which then
represents the short-distance dynamics. This relationship is very useful in 
order to understand  the saturation mechanism underlying the chiral two-pion 
exchange, as we shall discuss. A further feature of our treatment is that the 
properties of isospin symmetric nuclear matter (saturation point and 
compressibility) and of isospin asymmetric nuclear matter (asymmetry energy and
equation of state of pure neutron matter) are given in terms of only $one$ 
adjustable parameter, namely the cut-off $\Lambda$. It is then highly 
non-trivial to satisfy all known empirical constraints by fine-tuning just this
one single scale $\Lambda$ to a physically sensible value. We demonstrate that 
this is indeed possible. The present chiral perturbation theory calculation
allows to systematically explore the role that chiral symmetry plays in the
nuclear matter problem. 
 
Our paper is organized as follows. In section 2 we perform the chiral (or
small momentum) expansion of the nuclear matter equation of state
up-to-and-including terms of order ${\cal O}(k_f^5)$. We present analytical
results for the energy per particle $\bar E(k_f)$ as given by $1\pi$-exchange,
iterated $1\pi$-exchange and irreducible $2\pi$-exchange diagrams. Then we
discuss the saturation mechanism and the properties of our nuclear matter 
equation of state. In section 3 we calculate the density dependent asymmetry 
energy $A(k_f)$ in the same framework and discuss the corresponding results. 
Section 4 is devoted to the equation of state of pure neutron matter as it 
follows from the same set of pion-exchange diagrams. Finally, section 5 ends 
with a summary and an outlook.

\bigskip

\bild{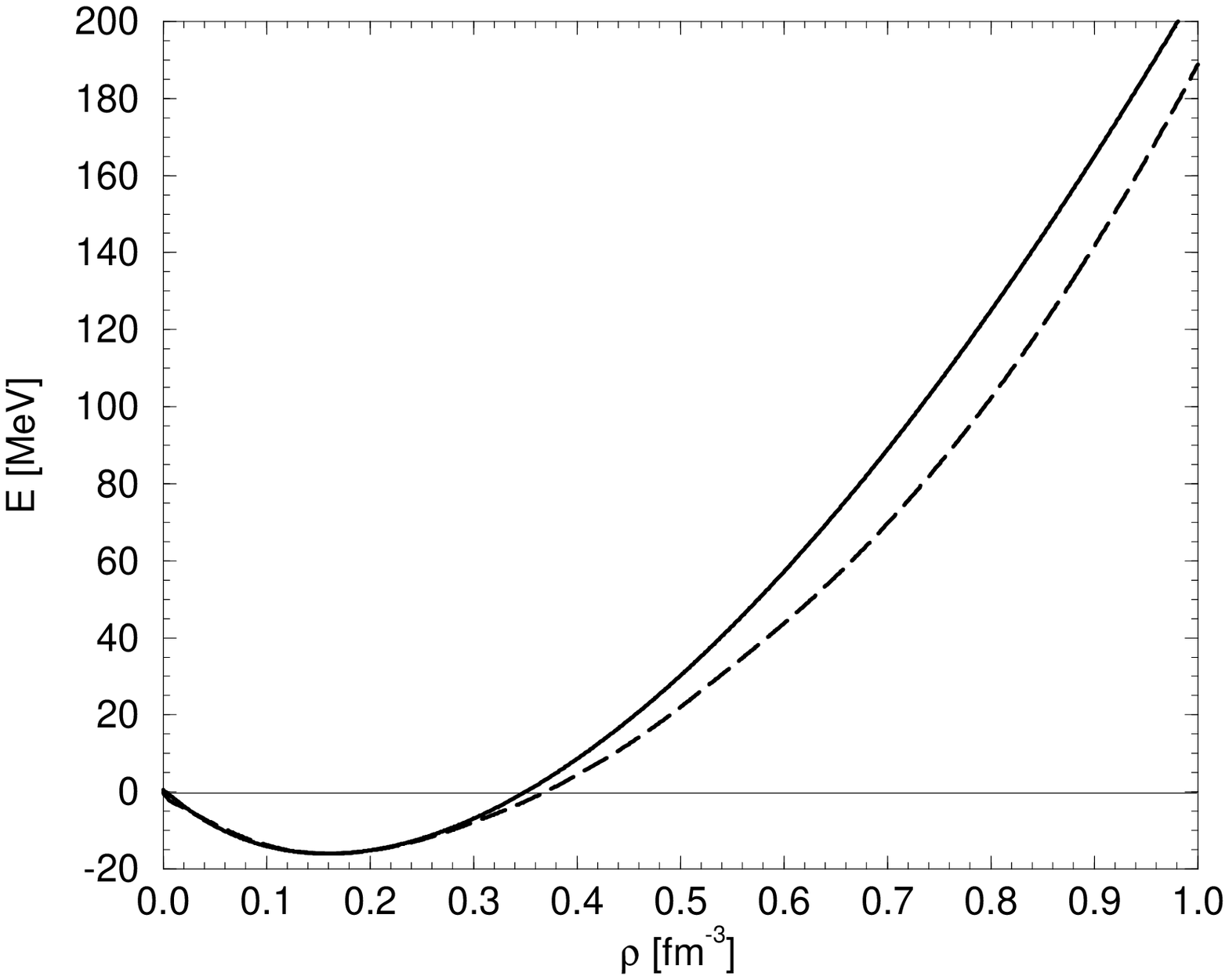}{12}
{\it Fig.\,1: Equation of state of isospin symmetric nuclear matter. The full 
line gives the parametrization eq.(1) with $\alpha = 5.27$ and $\beta=12.22$. 
The dashed line stems from the many-body calculation of ref.\cite{urbana}.} 

\bigskip 

Before presenting results of our (perturbative) diagrammatic calculation we 
like to draw attention to the following fact. A simple but nevertheless 
realistic parametrization of the energy per particle of isospin symmetric 
nuclear matter is given by the following expression,  
\begin{equation} \bar E(k_f) = {3k_f^2 \over 10 M} - \alpha\, {k_f^3 \over
M^2} + \beta \, {k_f^4 \over M^3} \,, \end{equation}
with $k_f$ the Fermi momentum related to the nucleon density in the usual way,
$\rho =2k_f^3/3\pi^2$, and $M=939$\,MeV the (free) nucleon mass. The first term
in eq.(1) is the well-known kinetic energy contribution of a Fermi gas and the 
powers of $M$ in the other terms have been chosen such that $\alpha$ and 
$\beta$ are dimensionless. The two-parameter form eq.(1) has generically a 
saturation minimum if $\alpha,\beta >0$. The interesting feature of this 
parametrization is that once $\alpha$ and $\beta $ are adjusted to the 
empirical nuclear matter saturation point, $k_{f0}=(1.35\pm0.05)\,$fm$^{-1}$ 
and $\bar E_0=\bar E(k_{f0})$, the nuclear matter compressibility
\cite{blaizot,vretenar}, 
\begin{equation} K= k_{f0}^2\, {\partial^2 \bar E(k_f) \over \partial k_f^2}
\bigg|_{k_f=k_{f0}} = (250\pm25)\,{\rm MeV} \,,\end{equation}
comes out correctly. For example, adjusting to the saturation point obtained 
in the sophisticated many-body calculation of the Urbana group \cite{urbana}, 
$\rho_0 = 2k_{f0}^3/3\pi^2 =0.159$\,fm$^{-3}$ and $\bar E_0=-16.0$\,MeV, one 
finds $\alpha =5.27$ and $\beta =12.22$, and the compressibility is $predicted$
as $K=236\,$MeV. In Fig.\,1 we compare the density dependence of $\bar E(k_f)$,
given by eq.(1) (full line) with the equation of state resulting from the
many-body calculation of ref.\cite{urbana} (dashed line). One observes
that the relative deviations of both curves do not exceed $10\%$ up to quite 
high densities, $\rho = 1.0$\,fm$^{-3}$. Keeping in mind that eq.(1) provides a
realistic parametrization of the nuclear matter equation of state will be very 
useful in order to understand the saturation mechanism underlying the chiral 
two-pion exchange. In the chiral limit ($m_\pi=0$) and truncating to order 
${\cal O}(k_f^4)$, the equation of state following from one- and two-pion 
exchange will be precisely of the form, eq.(1).

\section{Chiral expansion of the nuclear matter equation of state} 
The tool in order to systematically investigate the consequences of spontaneous
and explicit chiral symmetry breaking in QCD is chiral perturbation theory. 
Observables are calculated with the help of an effective field theory
formulated in terms of the Goldstone bosons (pions) and the low-lying baryons
(nucleons). The diagrammatic expansion in the number of loops has a one-to-one
correspondence to a systematic expansion of observables in small external 
momenta and meson (or quark) masses. For the problem under consideration, the 
external momentum is the Fermi momentum $k_f$, related to the nucleon density
$\rho=2k_f^3/3\pi^2$. The relevant observable is the 
energy per particle $\bar E(k_f)$, i.e. the ratio of energy density and 
particle density, with the free nucleon mass $M$ subtracted. Consequently, 
the equation of state of nuclear matter as given by chiral perturbation theory
will be of the form of an expansion in powers of the Fermi momentum  $k_f$. The
expansion coefficients are however non-trivial functions of $k_f/m_\pi$, the
dimensionless ratio of the two relevant small scales inherent to the
problem. Note that at the empirical saturation point, $k_{f0}  \simeq 2 m_\pi$,
so that both scales are of comparable magnitude for the densities of
interest. This also implies that pions have to be kept as explicit degrees of
freedom. At the saturation density pionic effects cannot be properly accounted
for just by coefficients of local NN-contact interactions which form the basis
of effective range expansions at finite density \cite{hammer}.  

Naive chiral power counting (which is basically a counting in terms of mass 
dimension) suggests that closed vacuum diagrams with $L$ loops (which represent
the ground state energy density in diagrammatic language) give rise to a 
contribution to the energy per particle of the form $\bar E(k_f) = k_f^{2L-1}
{\cal F}_L(k_f/m_\pi)$. However, the two-nucleon system is known to provide 
exceptions to the naive counting rules by the so-called iterated one-pion
exchange. In that case the relevant energy denominator is a difference of
nucleon kinetic energies, and this implies that the large scale factor $M$ (the
nucleon mass) appears in the numerator of the Feynman amplitude \cite{nnpap1}. 
There are indeed certain closed  three-loop diagrams (see Fig.\,3) which
contribute to the energy per particle already at order ${\cal O}(k_f^4)$ 
\cite{lutz}. By the same argument one expects that the four-loop
Fock-diagram\footnote{Obtained by drawing in the right diagram in Fig.\,3, a 
third pion-line intersecting both the other two.} which includes the 
twice-iterated one-pion exchange (proportional to $M^2$) will contribute 
already at order ${\cal O}(k_f^5)$. Luckily, the analogous  four-loop 
Hartree-diagram\footnote{Obtained by drawing in the left diagram in Fig.\,3 a 
third parallel pion-line connecting both equally oriented nucleon rings.} 
vanishes when taking the spin- or Dirac-trace. Because of the enormous 
complexity of four-loop diagrams we restrict ourselves to
the computation of all contributions up-to-and-including order ${\cal O}(k_f^5)
$ as they are given by closed three-loop diagrams. 

The only new ingredient in performing calculations at finite nucleon density
(as compared to calculations of scattering processes in the vacuum) is the 
in-medium nucleon propagator. It expresses the fact that the ground-state of
the system has changed from an empty vacuum to a filled Fermi sea of nucleons. 
In its relativistic form this in-medium propagator of a nucleon with
four-momentum $p^\mu =(p_0 ,\vec p\,)$ reads    
\begin{equation} (p \!\!\!/+M) \bigg\{ {i \over p^2-M^2+i \epsilon} - 2\pi\,
\delta(p^2-M^2)\, \theta(p_0)\,\theta(k_f-|\vec p\,|)\bigg\} \,. \end{equation}
Note that eq.(3) splits additively into the vacuum nucleon propagator and a 
medium insertion (i.e. the piece involving an on-shell delta-function and 
step-functions). This allows to organize the diagrammatic calculation according
to the number of medium insertions. Diagrams with no medium insertion lead to 
an unobservable shift of the vacuum energy. Diagrams with exactly one medium 
insertion just renormalize the nucleon mass to its measured value $M$ (because
of the presence of one single on-shell delta-function). The really interesting 
many-body effects from interactions start thus with diagrams having two or 
more medium insertions. 

The pion-nucleon interaction vertices relevant in this work are the 
pseudovector $\pi NN$-vertex and the Tomozawa-Weinberg $\pi\pi NN$-contact 
vertex of the form,  
\begin{equation} {g_A\over 2f_\pi}\, q\!\!\!/_a \gamma_5\, \tau_a \,, \qquad 
\qquad {1\over 4f_\pi^2}\, (q\!\!\!/_b-q\!\!\!/_a) \, \epsilon_{abc}\tau_c
\,. \end{equation}
Here, the pion four-momenta $q_{a,b}$ are out-going ones and $f_\pi=92.4\,$MeV
denotes the weak pion decay constant. For the nucleon axial vector coupling 
constant $g_A$ we choose the value $g_A=1.3$. Via the Goldberger-Treiman
relation this corresponds to a $\pi NN$-coupling constant of $g_{\pi N}=
g_A M/f_\pi=13.2$ which agrees with present empirical determinations of 
$g_{\pi N}$ from $\pi N$-dispersion relation analyses \cite{pavan}.

\subsection{Kinetic energy}
The first contribution to the energy per particle $\bar E(k_f)$ is the kinetic
energy of a non-interacting relativistic Fermi gas of nucleons. Expanding
$\sqrt{M^2+\vec p\,^2}-M$ in powers of $1/M$ and integrating over a 
Fermi sphere of radius $k_f$ one gets,    
\begin{equation} \bar E_k(k_f) = {3 k_f^2\over 10 M}-{3 k_f^4\over 56 
M^3}\,. \end{equation}
For the densities $\rho=2k_f^3/3\pi^2$ of interest this series converges very 
rapidly. The next term in this series, $k_f^6/48 M^5$, is already negligibly
small.    
\subsection{One-pion exchange}
\begin{center}

\SetWidth{2.5}
  \begin{picture}(400,100)

\ArrowArc(200,50)(40,0,180)
\ArrowArc(200,50)(40,180,360)
\DashLine(160,50)(240,50){6}
\Vertex(160,50){4}
\Vertex(240,50){4}

\end{picture}
\end{center}
{\it Fig.2: The one-pion exchange Fock-diagram. Solid and dashed lines 
represent nucleons and pions, respectively. The combinatoric factor of this 
diagram is 1/2.}

\vskip 0.7cm

The one-pion exchange Hartree-diagram is trivially zero (since $q^\mu = 0$) and
the one-pion exchange Fock-diagram is shown in Fig.\,2. The interesting
many-body effect comes from the medium insertion at each nucleon propagator. In
that case the two loop-integrations convert into integrals over Fermi spheres
of radius $k_f$. After taking the Dirac-trace of the product of $(p \!\!\!/+M)
$-factors and pseudovector $\pi N$-interaction vertices one performs a 
(non-relativistic) $1/M$-expansion of the complete integrand. In this form all 
integrals can be solved analytically and the result for the $1\pi$-exchange 
Fock-diagram reads,
\begin{eqnarray} \bar E_1(k_f) &=& {3g_A^2m_\pi^3 \over(4\pi f_\pi)^2} 
\bigg\{{u^3\over 3} +{1\over 8u} -{3u\over 4}+\arctan 2u -\Big( {3\over
8u}+{1\over 32u^3}\Big) \ln(1+4u^2) \nonumber \\&&+{m_\pi^2\over 40M^2} 
\bigg[{40\over 3}u^3-8u^5 +9u +{1\over 2u} -(12u^2+5)\arctan 2u -{1\over
8u^3}\ln(1+4u^2)\bigg] \bigg\}\,, \end{eqnarray}
Here, we have introduced the abbreviation $u=k_f/m_\pi$. The terms proportional
to $u^3$ and $u^5 m_\pi^2/M^2$ in eq.(6) correspond to the (repulsive) 
zero-range contact interaction produced by any pseudoscalar meson exchange. The
$1/M$-expansion is again converging rapidly. For example, at $k_f=270\,$MeV the
$1/M^2$-correction (second line in eq.(6)) is only $-5.2\%$ of the leading 
order term (first line in eq.(6)). Note also, the Taylor-series expansion of 
$\bar E_1(k_f)$ in eq.(6) converges only for $k_f \leq m_\pi/2$. This would
correspond to tiny densities of $\rho \leq 0.0027\,$fm$^{-3}$.      

\subsection{Iterated one-pion exchange}

\begin{center}

\SetWidth{2.5}
  \begin{picture}(400,100)

\ArrowArc(80,50)(40,90,270)
\ArrowArc(80,50)(40,-90,90)
\DashLine(80,90)(190,90){6}
\Vertex(80,90){4}
\Vertex(190,90){4}
\ArrowArc(190,50)(40,-90,-270)
\ArrowArc(190,50)(40,90,-90)
\DashLine(80,10)(190,10){6}
\Vertex(80,10){4}
\Vertex(190,10){4}

\ArrowArc(320,50)(40,0,180)
\ArrowArc(320,50)(40,180,360)
\DashLine(348.3,78.3)(291.7,21.7){6}
\DashLine(348.3,21.7)(291.7,78.3){6}
\Vertex(348.3,78.3){4}
\Vertex(291.7,21.7){4}
\Vertex(348.3,21.7){4}
\Vertex(291.7,78.3){4}

\end{picture}
\end{center}
{\it Fig.3: Iterated one-pion exchange Hartree and Fock-diagrams. The 
combinatoric factor of these diagrams is 1/4.}

\vskip 0.7cm

Next, we consider the three-loop diagrams shown in Fig.\,3. With two medium 
insertions on neighboring nucleon propagators they contribute, via mass and 
coupling constant renormalization, to the one-pion exchange Fock-diagram. These
effects are automatically taken care of in eq.(6) by using the physical input
parameters. Considering two medium insertions on non-neighboring nucleon 
propagators we encounter the planar box graph \cite{nnpap1}. In a
non-relativistic $1/M$-expansion the first contribution from the planar box
graph is the iterated $1\pi$-exchange which is enhanced  by the large scale
factor $M$. In case of the left diagram in Fig.\,3 (Hartree-diagram) the inner
loop integral becomes equal to the iterated $1\pi$-exchange amplitude in
forward direction. Using the analytical results given in section 4.3 of
ref.\cite{nnpap1}, we can even perform the remaining integral over the two
Fermi spheres of radius $k_f$. Altogether, we find for the Hartree-diagram in
Fig.\,3  with two medium insertions, 
\begin{equation} \bar E_2(k_f) = {3g_A^4Mm_\pi^4 \over 5(8\pi)^3f_\pi^4}\bigg\{
{9\over 2u}-59 u+(60+32u^2) \arctan 2u -\Big( {9\over 8u^3} +{35\over 2u}
\Big) \ln(1+4u^2) \bigg\} \,. \end{equation}  
Note that this expression does not include the contribution of a linear
divergence $\int_0^\infty dl\,1$ of the momentum space loop integral occurring
in the Hartree-diagram. In dimensional regularization such a linear divergence 
is set to zero, whereas in cut-off regularization it equals the momentum space
cut-off $\Lambda$. There is an interpretational problem with the result eq.(7)
when using dimensional regularization. While eq.(7) alone gives sizeable 
repulsion, for example $\bar E_2(k_f=270\,{\rm MeV })= 50.0\,$MeV, second order
perturbation theory arguments suggest attraction (because the 
intermediate-states lie mostly higher in energy than the initial or final 
state). The source of this apparent contradiction is the linear divergence 
$\int_0^\infty dl\,1$. Using cut-off regularization together
with a large enough value of $\Lambda$ one regains the expected attraction. The
additional additive terms specific for cut-off regularization are collected in 
eq.(15). The evaluation of the right diagram in Fig.\,3 (Fock-diagram) with 
two medium insertions is very similar. There one encounters the iterated
$1\pi$-exchange amplitude in backward direction \cite{nnpap1}. Putting  all 
pieces together one obtains the following result for the Fock-diagram in 
Fig.\,3 with two medium insertions, 
\begin{equation} \bar E_3(k_f) = {g_A^4 M m_\pi^4 \over (4\pi)^3f_\pi^4}\bigg\{
{u^3\over 2} + \int_0^u \!dx {3x (u-x)^2(2u+x) \over 2u^3(1+2x^2)} \Big[
(1+8x^2+8x^4) \arctan x-(1+4x^2)\arctan2x\Big] \bigg\}\,, \end{equation}
where we have again transferred the linear divergence proportional to the 
cut-off $\Lambda$ to eq.(15).

Next, we consider the diagrams in Fig.\,3 with three medium insertions. In this
case one has to evaluate an integral over the product of three Fermi spheres of
radius $k_f$. Suitable techniques allow to perform most of these nine 
integrations analytically. For the Hartree-diagram in Fig.\,3 with three
medium insertions we end up with the following representation,
\begin{equation} \bar E_4(k_f) = {9g_A^4 M m_\pi^4 \over (4\pi f_\pi)^4 u^3}
\int_0^u \!dx x^2\int_{-1}^1 \!dy \Big[2uxy+(u^2-x^2y^2)H\Big]\bigg\{
{2s^2+s^4\over 1+s^2}-2\ln(1+s^2)\bigg\}\,, \end{equation}
\begin{equation}H=\ln{u+xy\over u-xy}\,,\qquad s= xy+\sqrt{u^2-x^2+x^2y^2}\,.
 \end{equation} 
Similarly, we find for the Fock-diagram in Fig.\,3 with three medium 
insertions the representation,
\begin{equation} \bar E_5(k_f) = {9g_A^4 M m_\pi^4 \over (4\pi f_\pi)^4 u^3}
\int_0^u \!dx \bigg\{ {G^2\over 8}+{x^2\over 4}\int_{-1}^1\!dy \int_{-1}^1
\!dz {yz \,\theta(y^2+z^2-1) \over |yz|\sqrt{y^2+z^2-1}}\Big[s^2-\ln(1+s^2)
\Big] \Big[ \ln(1+t^2)-t^2\Big]\bigg\}\,, \end{equation}
with the auxiliary functions,
\begin{equation} G = u(1+u^2+x^2) -{1\over 4x}\big[1+(u+x)^2\big] \big[1+
(u-x)^2\big] \ln{1+(u+x)^2\over 1+(u-x)^2 } \,,  \end{equation} 
\begin{equation} t= xz+\sqrt{u^2-x^2+x^2z^2}\,. \end{equation}
For the numerical evaluation of the $dydz$-double integral in eq.(11) it is
advantageous to first anti-symmetrize the integrand both in $y$ and $z$ and
then to substitute $z= \sqrt{y^2\zeta^2 +1-y^2}$. This way the integration
region becomes equal to the unit-square $0\leq y,\zeta\leq1$. We note that our 
results, eqs.(7,8,9,11), which comprise all $2\pi$-exchange contributions at
order ${\cal O}(k_f^4)$, agree with those of ref.\cite{lutz} after suitable 
simplification. As a further check on the whole formalism we rederived the 
analytical result of  Onsager et al. \cite{onsager} for the two-photon 
exchange Fock-diagram of the electron-gas, $\bar E= [(\ln2/3)-(3\zeta(3)/2\pi^2
)]\,Rydberg$, using the same techniques.

The diagrams in Fig.\,3 with four medium insertions do not have to be 
calculated explicitly. These are purely imaginary and they are cancelled by the
imaginary parts coming along with the diagrams with two and three medium 
insertions. Via unitarity, such imaginary parts are related to on-shell 
scattering of nucleons. Energy conservation and Pauli-blocking in the medium
leave no phase-space for these processes and thus the energy per particle 
$\bar E(k_f)$ remains a real  quantity (as it must of course be).

\subsection{Irreducible two-pion exchange}

\begin{center}

\SetWidth{2.5}
  \begin{picture}(400,100)

\ArrowArc(50,50)(40,0,180)
\ArrowArc(50,50)(40,180,360)
\DashCurve{(10,50)(50,70)(90,50)}{6}
\DashCurve{(10,50)(50,30)(90,50)}{6}
\Vertex(10,50){4}
\Vertex(90,50){4}

\ArrowArc(150,50)(40,0,180)
\ArrowArc(150,50)(40,180,360)
\DashLine(110,50)(178.3,78.3){6}
\DashLine(110,50)(178.3,21.7){6}
\Vertex(178.3,78.3){4}
\Vertex(110,50){4}
\Vertex(178.3,21.7){4}

\ArrowArc(250,50)(40,0,180)
\ArrowArc(250,50)(40,180,360)
\DashLine(287,65)(213,65){6}
\DashLine(213,35)(287,35){6}
\Vertex(287,65){4}
\Vertex(287,35){4}
\Vertex(213,65){4}
\Vertex(213,35){4}

\ArrowArc(350,50)(40,0,180)
\ArrowArc(350,50)(40,180,360)
\DashLine(378.3,78.3)(321.7,21.7){6}
\DashLine(378.3,21.7)(321.7,78.3){6}
\Vertex(378.3,78.3){4}
\Vertex(321.7,21.7){4}
\Vertex(378.3,21.7){4}
\Vertex(321.7,78.3){4}

\end{picture}
\end{center}
{\it Fig.4: Irreducible two-pion exchange Fock-diagrams. The corresponding
Hartree-diagrams vanish for isospin symmetric nuclear matter. The combinatoric 
factors of these diagrams are 1/4, 1, 1/2 and 1/4, in the order shown.}

\vskip 0.7cm
Contributions of three-loop $2\pi$-exchange diagrams to $\bar E(k_f)$ at order 
${\cal O}(k_f^5)$ are in accordance with the naive chiral power counting rules.
The relevant $2\pi$-exchange Fock-diagrams are shown in Fig.\,4. Considering 
two medium insertions at the upper and lower nucleon propagators one finds that
their contribution to the energy per particle can be expressed in terms of the
T-matrix related to the irreducible $2\pi$-exchange. This T-matrix has been
evaluated in section 4.2 of ref.\cite{nnpap1} using dimensional
regularization. The result of cut-off regularization differs only little by
some coefficients of purely polynomial terms and the presence of a quadratic
divergence $\Lambda^2$ (see also appendix\,A in ref.\cite{epel}). After 
performing the remaining integral over two Fermi spheres of radius $k_f$ one 
ends up with the following analytical expression for the Fock-diagrams in 
Fig.\,4, 
\begin{eqnarray} \bar E_6(k_f) &=& {m_\pi^5 \over (4 \pi f_\pi)^4}\bigg\{\bigg[
{3\over 32u^3}( 43g_A^4+6g_A^2 -1) +{3\over 4u} (23g_A^4+2g_A^2-1) \bigg] \, 
\ln^2(u+\sqrt{1+u^2}) \nonumber \\ && +\bigg[{u^4\over 5} (11g_A^4 -10 g_A^2
-1) +{u^2\over 10} (59g_A^4-50g_A^2-9)+{1\over 40} (883g_A^4-90g_A^2 -73) 
\nonumber \\ && +{3\over16u^2}(1-6g_A^2-43g_A^4)\bigg] \sqrt{1+u^2} \ln(u+\sqrt
{1+u^2}) +{3\over32 u} (43g_A^4+6g_A^2-1)\nonumber \\ && +{u\over 160} (397+210
g_A^2-5647 g_A^4) + {u^3\over 5} (4+5g_A^2+31g_A^4)+{u^5\over 600}(119+710
g_A^2 -349 g_A^4) \nonumber \\ && +  \Big[ u^3(15g_A^4-6g_A^2-1) +{u^5 \over
5} (11g_A^4-10g_A^2-1)\Big] \ln{m_\pi \over 2\Lambda} \bigg\}\,,\end{eqnarray} 
where we have again transferred the term quadratic in the cut-off $\Lambda$ to
eq.(15). Obviously, terms proportional to $g_A^0$, $g_A^2$, $g_A^4$ in
eq.(14) belong to the first, second, third and fourth diagram in Fig.\,4,
respectively. Note that the last diagram contributes both at order 
${\cal O}(k_f^4)$ and at order ${\cal O}(k_f^5)$. The contribution of the 
Hartree-diagrams related to irreducible $2\pi$-exchange vanishes in isospin 
symmetric nuclear matter for the following simple reason. After taking spin- 
and isospin-traces of the T-matrix, only the isoscalar central NN-amplitude 
$V_C(0)$ survives, and this amplitude receives no contribution from irreducible
chiral $2\pi$-exchange at leading order as shown in section 4.2 of 
ref.\cite{nnpap1}. The diagrams in Fig.\,4 with three medium insertions 
contribute only via relativistic $1/M$-corrections at order 
${\cal O}(k_f^6)$. Besides eq.(14), there is no further contribution at
order ${\cal O}(k_f^5)$ from three-loop $2\pi$-exchange diagrams. Note also
that all contributions to $\bar E(k_f)$ remain finite in the chiral limit
$m_\pi=0$. For example, eq.(14) becomes proportional to $k_f^5[\ln(k_f/\Lambda)
+const]$. As far as we can see, there are no infra-red singularities
generated by the Goldstone bosons (massless pions). 

Finally, we give the complete expression for the power divergences specific to
cut-off regularization,   
\begin{equation} \bar E_\Lambda(k_f) = {\Lambda \,k_f^3 \over (4\pi f_\pi)^4} 
\Big[ -10 g_A^4 M +(3g_A^2+1)(g_A^2-1) \Lambda \Big] \,. \end{equation}
Obviously, the first term proportional to $M$ stems from iterated 
$1\pi$-exchange. The Hartree and Fock-diagrams in Fig.\,3 contribute to it 
in the ratio $4:1$ as a result of their spin- and isospin factors. The term 
quadratic in the cut-off $\Lambda$, on the other hand, originates from 
irreducible $2\pi$-exchange.

Eq.(15) is equivalent to the contribution of a zero-range NN-contact 
interaction. The cut-off scale $\Lambda$ is effectively parametrizing its 
strength. Note that eq.(15) leads to a sizeable attraction in the hundred MeV
range for physically reasonable values of the cutoff, $0.5\,$GeV$\,<\Lambda
<1.0\,$GeV. We are thus lead to the conclusion that in isospin symmetric
nuclear matter $2\pi$-exchange produces "zero-range" attraction and 
finite-range repulsion.  It should also be noted that the dominant (real part 
of) iterated $1\pi$-exchange does not have an equivalent representation as a
local  coordinate-space NN-potential (see section 4.3 in ref.\cite{nnpap1}). 
Such a connection exists only for $1\pi$-exchange and (with certain 
restrictions concerning the zero-range part) for irreducible $2\pi$-exchange 
\cite{nnpap1}. Note that $\bar E_\Lambda(k_f)$ in eq.(15) encodes the
short-distance NN-dynamics necessary for nuclear binding. We emphasize that,
for reasons of consistency, this piece (which originates from the few divergent
parts of chiral $2\pi$-exchange) should $not$ by further iterated either with
itself or with $1\pi$-exchange, in contrast to what has been done with the
NN-contact interaction in ref.\cite{lutz}.

\subsection{Results}
Before presenting results for the nuclear matter equation of state, we would 
like to exhibit the saturation mechanism. For that purpose we truncate the 
previous calculation of one- and two-pion exchange diagrams at order ${\cal O}
(k_f^4)$ and we take the chiral limit, $m_\pi=0$, (disregarding for the moment 
the less important $\Lambda^2$-term in eq.(15)). In that case the complete
expression for the energy per particle $\bar E(k_f)$  turns into the form, 
eq.(1),  with coefficients $\alpha$ and $\beta$ given by
\begin{equation} \alpha =10 \Big({g_{\pi N}\over 4\pi}\Big)^4  
{\Lambda \over M}- \Big({g_{\pi N}\over 4\pi} \Big)^2  \,, \end{equation}
\begin{equation} \beta = {3\over 70} \Big({g_{\pi N}\over 4\pi}\Big)^4(4\pi^2+
237-24\ln2)-{3\over 56} =13.55 \,. \end{equation}  
Note that the parameterfree expression for $\beta$ gives a number quite close 
to $\beta =12.22$ as extracted from the realistic equation of the state of 
ref.\cite{urbana}. Furthermore, by adjusting the cut-off scale to $\Lambda=(0.5
\dots 0.6)M$ (which in fact lies in the physically reasonable range) $\alpha$
will take on its required value. Therefore, as long as the effects due to the
finite pion mass and the terms at order ${\cal O}(k_f^5)$ do not change this
picture completely, realistic nuclear binding is guaranteed by chiral
pion-nucleon dynamics, together with fine-tuning of the single scale $\Lambda$ 
representing the short-distance NN-dynamics.

Now we include all calculated terms up to order ${\cal O}(k_f^5)$ and we set
$m_\pi=135\,$MeV (the neutral pion mass). We fix the (negative) binding energy
to the value $\bar E_0=-15.26\,$MeV obtained in extensive and elaborate fits of
nuclide masses in ref.\cite{atommass}. The energy per particle $\bar E(k_f)$ 
has a minimum with this value of $\bar E_0$ at a density of $\rho_0 = 0.178\,
$fm$^{-3}$ (corresponding to a Fermi momentum of $k_{f0}= 272.7\,{\rm MeV}
=1.382\,$fm$^{-1}$) if the cut-off scale is fine-tuned to $\Lambda = 646.3\,
$MeV.  Note that a saturation density of $\rho_0=0.178\,$fm$^{-3}$ is somewhat 
on the large side. A slight increase of $g_A$ by $2\%$ would even shift $\rho_0
$ to the center of the empirical saturation density $\rho_0=0.166\,$fm$^{-3}$ 
\cite{blaizot}. Let us also investigate how the value $\bar E_0=-15.26\,$MeV 
arises in the present calculation. Its decomposition into contributions from 
the kinetic energy and the three classes of diagrams is $\bar E_0=(23.40+18.24
-68.35+11.45)\,$MeV. On the other hand, when ordering in chiral powers ${\cal 
O}(k_f^\nu), (\nu=2,3,4,5)$ one finds $\bar E_0 = (23.75 -154.54+124.61-9.08)\,
$MeV. The individual entries follow closely the pattern prescribed by the
parametrization eq.(1) with a balance between large third and fourth order
terms. If the behavior of the last two entries is representative the chiral
expansion of $\bar E_0$ appears to converge.   

\bigskip

\bigskip

\bild{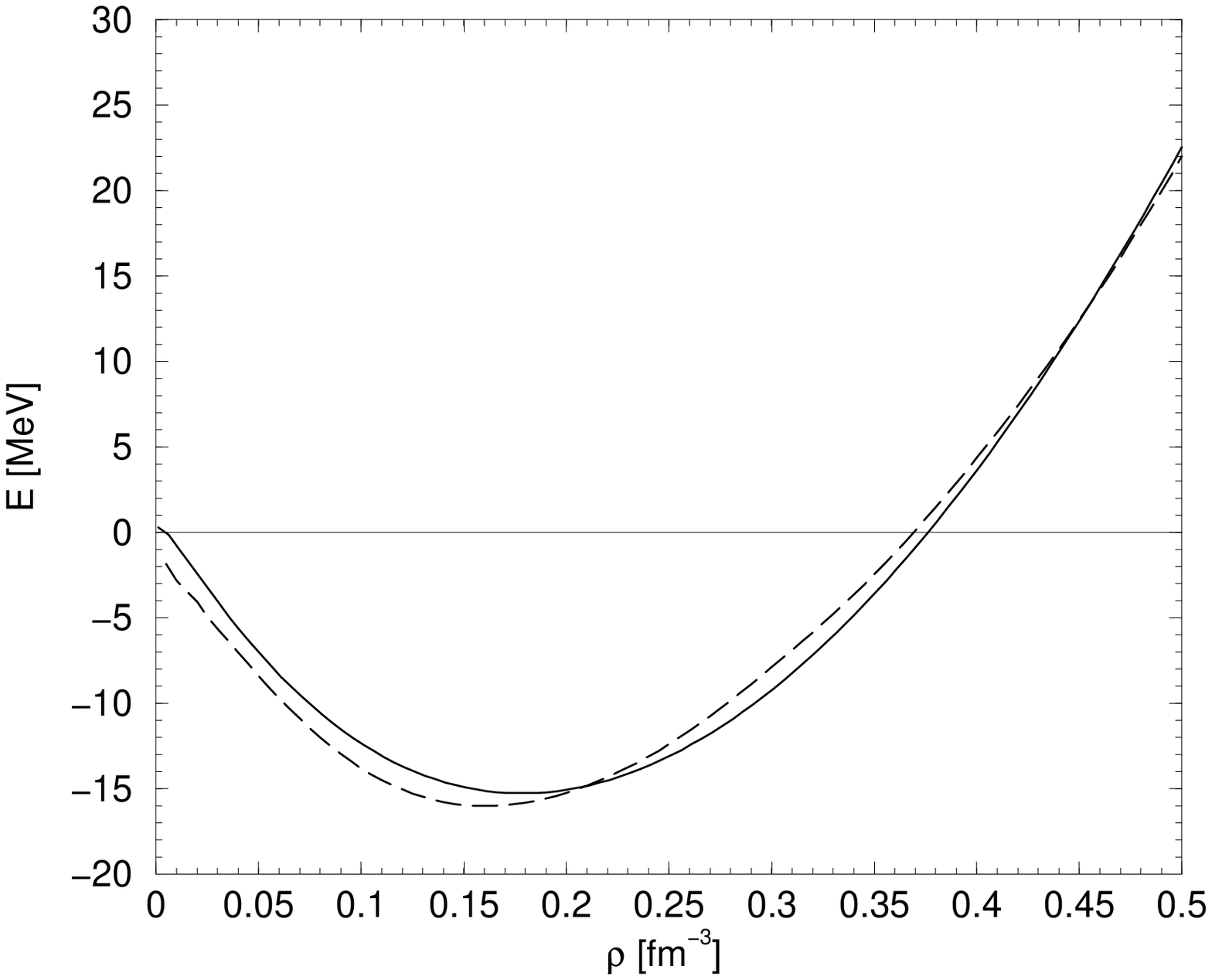}{10}
{\it Fig.\,5: Energy per particle of isospin symmetric nuclear matter derived 
from chiral one-and two-pion exchange (solid line). The value of the cut-off 
scale is $\Lambda = 646\,$MeV. The dashed line corresponds to the result of
ref.\cite{urbana}.}

\bigskip

In Fig.\,5 we show the resulting nuclear equation of state for densities up to
$\rho=0.5\,$fm$^{-3}$ (i.e. about $3\rho_0$). The nuclear compressibility $K$
related to the curvature of the saturation curve at its minimum
comes out as $K=255\,$MeV, in very good agreement with the nowadays accepted 
empirical value $K=(250\pm25)\,$MeV \cite{blaizot,vretenar} (for the definition
of $K$, see eq.(2)).

Fig.\,6 shows by the solid line the dependence of the saturation point 
$(\rho_0, \bar E_0)$ on the cut-off $\Lambda$ which has been varied in the 
range $0.6\,{\rm GeV} < \Lambda < 0.7\,$GeV. The inserted rectangle corresponds
to the empirical saturation ``point'' $\bar E_0 = (-16\pm 1)$\,MeV  and 
$k_{f0}=(1.35\pm0.05)\,$fm$^{-1}$ quoted in ref.\cite{rolf}. The variation 
of the saturation point $(\rho_0,\bar E_0)$ in Fig.\,6 is somewhat reminiscent
of the familiar Coester-line; of course, the physics behind it is quite 
different here. It is gratifying that our line meets the empirical 
saturation  ``point''. The dashed line in Fig.\,6, which gives the calculated 
saturation point in the chiral limit, $m_\pi=0$ (keeping $g_A,M,f_\pi$ fixed), 
is also interesting. It suggests that in QCD explicit chiral symmetry breaking
is not a crucial condition for nuclear binding. Note however that for fixed
$\Lambda$, taking the chiral limit $m_\pi =0$, increases the binding energy
$\bar E_0$ as well as the equilibrium density $\rho_0$ considerably.    

\bigskip

\bigskip

\bild{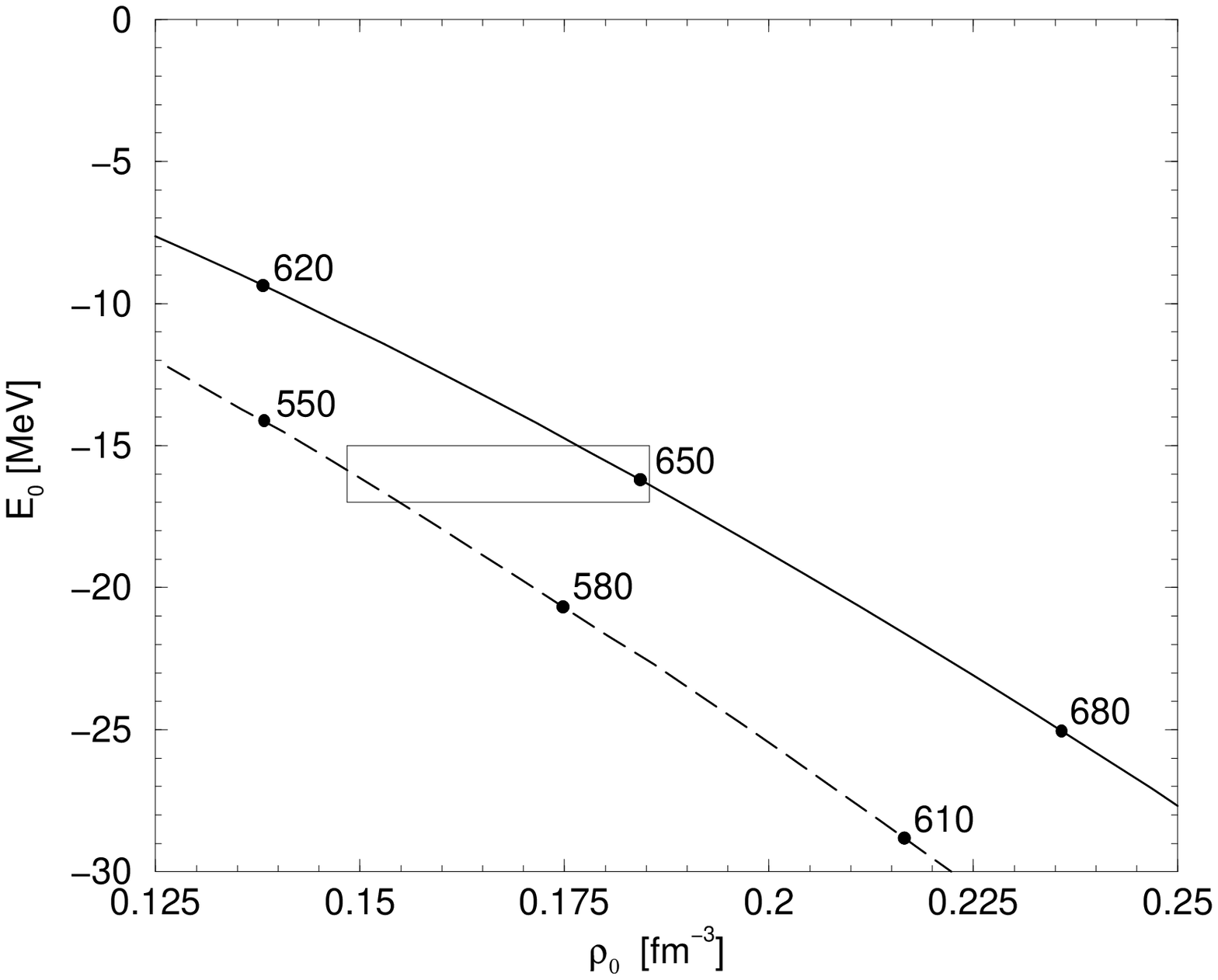}{10}
{\it Fig.\,6: The nuclear matter saturation point $(\rho_0,\bar E_0)$ at 
finite pion mass (solid line) and in the chiral limit (dashed line) as a 
function of the cut-off scale $\Lambda$ (points in MeV). The inserted rectangle
corresponds to the empirical saturation ``point''.}

\bigskip
   
\section{Asymmetry energy}
The properties of isospin symmetric matter can be well reproduced in our
framework with just one single adjustable cut-off scale $\Lambda$. In order to
test the isospin dependence of the underlying pion-exchange dynamics which
causes saturation, we now investigate the asymmetry energy. In isospin 
asymmetric matter the Fermi seas of protons and neutrons are filled unequally. 
With the help of the projection operators $(1\pm \tau_3)/2$ such an
isospin-asymmetric situation is realized by the simple substitution, 
\begin{equation} \theta(k_f-|\vec p\,|) \quad \to \quad {1+\tau_3 \over 2}\,  
\theta(k_p-|\vec p\,|) +{1-\tau_3 \over 2}\, \theta(k_n-|\vec p\,|)\,,
\end{equation}
in the in-medium nucleon propagator eq.(3). Here $k_p$ and $k_n$ denote the 
(different) Fermi momenta of protons and neutrons. Choosing $k_{p,n} = k_f(1
\mp \delta)^{1/3}$ (with $\delta$ a small parameter) the nucleon density 
$\rho=\rho_p+\rho_n= (k_p^3+k_n^3)/3\pi^2= 2k_f^3/3\pi^2$ stays constant. The 
expansion of the energy per particle of isospin asymmetric  nuclear matter,  
\begin{equation} \bar E_{as}(k_p,k_n)= \bar E(k_f) + \delta^2\, A(k_f)+ \dots  
\,, \end{equation}
around the symmetry line ($k_p=k_n$ or $\delta=0$) defines the asymmetry energy
$A(k_f)$. Note that the parameter $\delta$ is equal to $(\rho_n-\rho_p)/
(\rho_n+\rho_p)$ or $(N-Z)/(N+Z)$. Differences in comparison with the 
diagrammatic calculation in section 2 occur only with respect to isospin 
factors and the radii of the Fermi spheres, $k_p = k_f(1- \delta)^{1/3}$ or
$k_n = k_f(1+ \delta)^{1/3}$. Following the scheme in section 2, we summarize
the individual contributions  to the asymmetry energy $A(k_f)$ without going
into further technical details.  

\medskip

\noindent i) Kinetic energy of a relativistic Fermi gas:
\begin{equation} A_k(k_f)={k_f^2\over 6 M}-{k_f^4\over 12 M^3}\,.\end{equation}
The next term in this series, $k_f^6/16 M^5$, is negligibly small.

\medskip

\noindent ii) $1\pi$-exchange Fock-diagram including the relativistic
$1/M^2$-correction: 
\begin{equation} A_1(k_f) = {g_A^2m_\pi^3 \over(4\pi f_\pi)^2} 
\bigg\{\Big( {u\over 3}+{1\over 8u}\Big) \ln(1+4u^2)-{u\over 2}-{u^3\over 3} 
+{m_\pi^2\over M^2} \bigg[u^3-{u^2\over 2} \arctan 2u -{u^3
\over 3}\ln(1+4u^2)\bigg] \bigg\}\,, \end{equation}
with the abbreviation $u=k_f/m_\pi$.

\medskip

\noindent iii) Hartree-diagram in Fig.\,3 with two medium insertions: 
\begin{equation} A_2(k_f)= {g_A^4 M m_\pi^4 \over (8\pi)^3f_\pi^4}\bigg\{\Big( 
{25\over 3}u +{7\over 6u}\Big) \ln(1+4u^2)-{14\over 3} u-16u^2 \arctan 2u
\bigg\} \,. \end{equation}

\noindent iv) Fock-diagram in Fig.\,3 with two medium insertions:
\begin{equation} A_3(k_f) = {g_A^4 M m_\pi^4 \over (4\pi)^3f_\pi^4}\bigg\{
-{5\over 6}u^3 + \int_0^u \!dx {3x^2-4u^2 \over 6u(1+2x^2)} \Big[
(1+8x^2+8x^4) \arctan x-(1+4x^2)\arctan2x\Big] \bigg\}\,. \end{equation} 

\medskip

\noindent v) Hartree-diagram in Fig.\,3 with three medium insertions:
\begin{eqnarray} A_4(k_f)&=&{g_A^4 M m_\pi^4 \over (4\pi f_\pi)^4 u^3} \int_0^u
\! dx x^2 \int_{-1}^1 \! dy \bigg\{ \bigg[{uxy(26u^2-30x^2y^2) \over 3
(u^2-x^2y^2)} +(3u^2-5x^2y^2)H\bigg] \nonumber \\ && \times \bigg[{2s^2+s^4
\over 1+s^2}-2\ln(1+s^2) \bigg]  -{4u^2 H \,s^6 \over 3(1+s^2)^2}+\Big[2uxy+
(u^2-x^2y^2) H\Big] \nonumber \\ && \times \Big[(5+s^2)(3s^2-8s s' 
+8 s'^2)+8s(1+s^2)(s''-5s'+3s)\Big] {s^4\over 3(1+s^2)^3} 
\bigg\} \,. \end{eqnarray}
The auxiliary functions $H$ and $s$ have been defined in eq.(10) and $s'$ and 
$s''$ denote partial derivatives,   
\begin{equation} s' = u {\partial s \over \partial u} \,, \qquad s'' = u^2 
{\partial^2 s \over \partial u^2} \,. \end{equation}

\medskip

\noindent vi) Fock-diagram in Fig.\,3 with three medium insertions:
\begin{eqnarray}  A_5(k_f) &=&{g_A^4Mm_\pi^4\over(4\pi f_\pi)^4 u^3}\int_0^u\!
dx\bigg\{{G\over 24}(3G_{20}-2G_{11}+3G_{02}-8G_{01}-3G) +{G_{10}+G_{01}\over
24}(3G_{10}-5G_{01}) \nonumber \\ && +{x^2\over 6}\int_{-1}^1\!dy \int_{-1}^1
\!dz {yz \,\theta(y^2+z^2-1) \over |yz|\sqrt{y^2+z^2-1}} \bigg[ {2s^3t^4(8s'
-3s) \over(1+s^2)(1+t^2)} \nonumber \\ &&+\Big[(3+s^2)(8ss'-3s^2-8s'^2)+4s
(1+s^2)(6s'-3s-2s'')\Big] {s^2[t^2-\ln(1+t^2)]\over (1+s^2)^2}\bigg] \bigg\}
\,.  \end{eqnarray}
The auxiliary functions $G$ and $t$ have been defined in eqs.(12,13) and we 
have introduced the following double-index notation for partial derivatives 
of the function $G$,
\begin{equation} G_{ij} := x^i u^j {\partial^{i+j}G \over \partial x^i \partial
u^j} \,, \quad 1\leq i+j \leq 2\,. \end{equation}

\medskip

\noindent vii)  Irreducible $2\pi$-exchange Hartree- and Fock-diagrams in
Fig.\,4: 
\begin{eqnarray} A_6(k_f) &=& {m_\pi^5 \over (4 \pi f_\pi)^4}\bigg\{ \bigg[
{1\over 12u} (1-2g_A^2-23g_A^4)+{u\over 3} (1+2g_A^2-7g_A^4)  \bigg] \, 
\ln^2(u+\sqrt{1+u^2}) \nonumber \\ && +\bigg[{1\over 6} (23g_A^4 +2 g_A^2-1) 
+u^2\Big({1\over 3}+2g_A^2-5g_A^4\Big)-{16\over 3}g_A^4 u^4\bigg] \sqrt{1+u^2} 
\ln(u+\sqrt {1+u^2}) \nonumber \\ && +{u\over 12} (1-2g_A^2-23 g_A^4) + u^3
\Big({1\over 4}+{11\over 2} g_A^2-{245\over 12}g_A^4\Big)-{u^5\over 27}(1+10
g_A^2 +g_A^4) \nonumber \\ && + u^3 \bigg[{5\over 3}+10g_A^2-25g_A^4 -{16 \over
3}g_A^4 u^2\bigg] \ln{m_\pi \over 2\Lambda} \bigg\}\,.\end{eqnarray}
The Hartree-diagrams (not shown in Fig.\,4) give a contribution to the
asymmetry energy $A(k_f)$ through the isovector central NN-amplitude $W_C(0)$ 
(see section 4.2 in ref.\cite{nnpap1}). In terms of this amplitude the 
contribution of the irreducible $2\pi$-exchange Hartree-diagrams reads 
$A(k_f) = -\rho\, W_C(0)/2$.   
 
\medskip

\noindent viii) Power divergences specific for cut-off regularization:
\begin{equation} A_\Lambda(k_f) = {\Lambda \,k_f^3 \over 3(4\pi f_\pi)^4}\Big[
26 g_A^4 M +5(3g_A^2+1)(1-g_A^2) \Lambda\Big]\,.\end{equation}
Again, this expression collects the contributions from iterated $1\pi$-exchange
and irreducible $2\pi$-exchange. In comparison to $\bar E_\Lambda(k_f)$ given 
in eq.(15) the term, eq.(29), has changed sign. The isospin-structure of the 
underlying $2\pi$-exchange interaction shows up in the linear and quadratic
term in $\Lambda$ through relative factors $-13/15$ and $-5/3$, respectively. 

As a side remark we note that in the chiral limit, $m_\pi=0$, the order 
${\cal O}(k_f^4)$ contributions to the asymmetry energy $A(k_f)$
(see eqs.(22,23,24,26)) can be evaluated in closed form with the result,  
\begin{equation} A(k_f)|_{m_\pi=0} = -\bigg( {g_A k_f\over 4 \pi f_\pi} 
\bigg)^4  {M\over 135} ( 44\pi^2 +987+216\ln 2) \,. \end{equation}

For the numerical evaluation of the asymmetry energy we use the same parameter 
input as in section 2.5, in particular a cut-off scale of $\Lambda=646.3\,$MeV.
The value of the asymmetry energy $A(k_f)$ at our saturation point $k_{f0} = 
272.7\,$MeV comes out as $A_0 =A(k_{f0}) =33.8\,$MeV. This is in very 
good agreement with the empirical value of $A_0 = 33.2\,$MeV obtained in 
ref.\cite{atommass} from extensive and elaborate fits to nuclide masses. 
Let us again investigate how the value $A_0=33.8\,$MeV arises in the present
calculation. Its decomposition into contributions from the kinetic energy and 
the three classes of diagrams is $A_0=(12.7-5.3+40.7-14.3)\,$MeV. On the 
other hand, when ordering in chiral powers ${\cal O}(k_f^\nu), (\nu=2,3,4,5)$ 
one finds $A_0 = (13.2 + 129.9 -127.4+18.1)\,$MeV. Here, one observes an 
almost complete cancelation of large third and fourth order terms. Again, if 
the behavior of the last two entries is representative, the chiral expansion of
the asymmetry energy $A_0$ should converge.

In Fig.\,7 we show by the solid line the dependence of the asymmetry energy
$A(k_f)$ on the density $\rho = 2k_f^3/3\pi^2$. One observes that $A(k_f)$ 
reaches its maximum close to the saturation density $\rho_0$. We obtain 
therefore a small value of the (empirically not well determined) slope 
parameter $L=k_{f0} [\partial A(k_f)/\partial k_f]_{0} = 14.7\,$MeV (see also 
ref.\cite{blaizot}).  Furthermore, we extract an asymmetry compressibility of 
$K_{as}=k^2_{f0} [\partial^2 A(k_f)/\partial k^2_f]_{0} -2L = -359\,$MeV. Such 
large and negative values of $K_{as}$ are also found in some other calculations
(see table\,4 in ref.\cite{blaizot}). For comparison we show in Fig.\,7 also
the result for the asymmetry energy $A(k_f)$ obtained in the 
Br\"uckner-Hartree-Fock calculation of ref.\cite{bombaci}. The downward 
bending of our asymmetry energy $A(k_f)$ at densities $\rho >0.2$\,fm$^{-3}$
presumably indicates the limits of validity of the present chiral perturbation
theory calculation of nuclear matter.

\bigskip

\bigskip

\bild{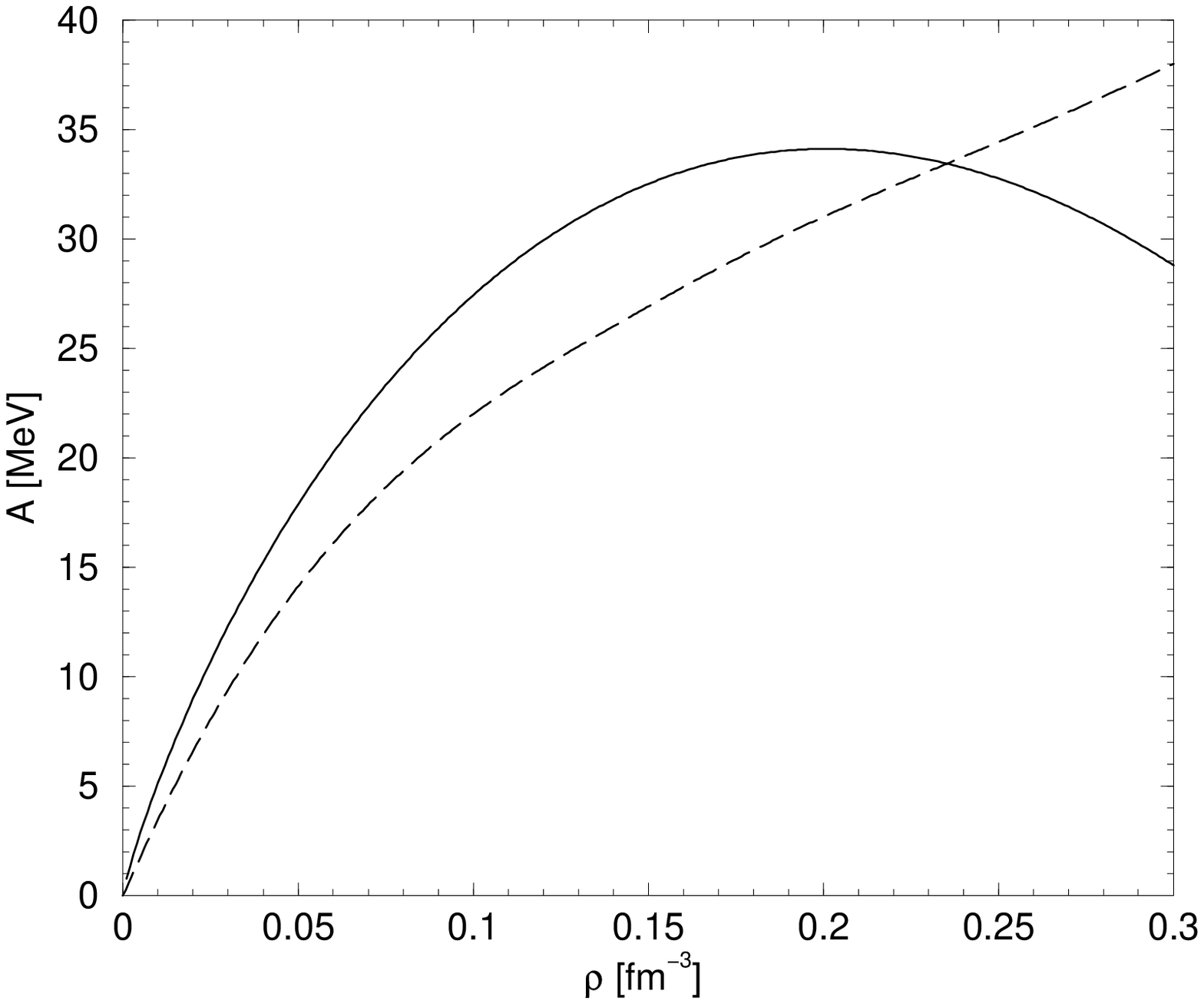}{11}
{\it Fig.\,7: The density dependence of the asymmetry energy $A(k_f)$ (solid
line). The cut-off is $\Lambda =646\,$MeV. The dashed line gives the result 
of ref.\cite{bombaci}.}

\bigskip

The simultaneous correct description of several nuclear matter properties 
(saturation point, compressibility, asymmetry energy) with only one adjustable 
scale parameter, the cut-off $\Lambda$, is indeed highly non-trivial. This 
strongly indicates that nuclear binding and saturation through the combination
of chiral one- and two-pion exchange and short-distance dynamics encoded in a
single high momentum scale $\Lambda$ is a realistic scenario (at low densities
$\rho \leq 0.2\,$fm$^{-3}$). Our treatment of the zero-range components of the 
effective NN-interaction in terms of a cut-off $\Lambda$ seems to work 
surprisingly well and it resembles effects of correlation functions 
in sophisticated many-body calculations.   

\section{Neutron matter}
The extreme of asymmetric nuclear matter is pure neutron matter. All existing 
realistic calculations \cite{urbana,wiringa,glend,lirolf} agree that pure 
neutron matter is unbound. Its energy per particle monotonically rises with the
neutron density.  

In order to arrive at the energy per particle, $\bar E_n(k_n)$, of neutron 
matter in our diagrammatic framework it is sufficient to make the substitution 
\begin{equation} \theta(k_f-|\vec p\,|) \quad \to \quad {1-\tau_3 \over 2}\, 
\theta(k_n-|\vec p\,|)\,,\end{equation}
in the in-medium nucleon propagator eq.(3). Here $k_n$ denotes the
Fermi momentum of the neutrons related to the neutron density by $\rho_n=k_n^3
/3\pi^2$. As a consequence of the substitution eq.(31) only the isospin factors
of individual diagrams change and all Fermi spheres have the radius $k_n$. We 
follow the scheme in section 3 and enumerate the individual contributions to
the energy per particle, $\bar E_n(k_n)$, of neutron matter.

\medskip

\noindent i) Kinetic energy of a relativistic Fermi gas:
\begin{equation} \bar E_{n,k}(k_n) = {3 k_n^2\over 10 M}-{3 k_n^4\over56M^3}
\,.  \end{equation}

\medskip

\noindent ii) $1\pi$-exchange Fock-diagram including the relativistic
$1/M^2$-correction: 

\begin{eqnarray} \bar E_{n,1}(k_n) &=& {g_A^2m_\pi^3 \over(4\pi f_\pi)^2} 
\bigg\{{u^3\over 3} +{1\over 8u} -{3u\over 4}+\arctan 2u -\Big( {3\over
8u}+{1\over 32u^3}\Big) \ln(1+4u^2) \nonumber \\&&+{m_\pi^2\over 40M^2} 
\bigg[{40\over 3}u^3-8u^5 +9u +{1\over 2u} -(12u^2+5)\arctan 2u -{1\over
8u^3}\ln(1+4u^2)\bigg] \bigg\}\,. \end{eqnarray}
We emphasize that in this section the meaning of $u$ changes to $u=k_n/m_\pi$. 

\medskip

\noindent iii) Hartree-diagram in Fig.\,3 with two medium insertions:

\begin{equation} \bar E_{n,2}(k_n) = {g_A^4 M m_\pi^4\over10(8\pi)^3f_\pi^4}
\bigg\{{9\over 2u} -59 u +(60+32u^2) \arctan 2u -\Big( {9\over 8u^3} +{35\over
2u} \Big) \ln(1+4u^2) \bigg\} \,. \end{equation}  

\medskip

\noindent iv) Fock-diagram in Fig.\,3 with two medium insertions:

\begin{equation} \bar E_{n,3}(k_n) = {g_A^4 M m_\pi^4 \over (4\pi)^3f_\pi^4}
\bigg\{-{u^3\over 6} + \int_0^u \!dx {x (u-x)^2(2u+x) \over 2u^3(1+2x^2)} \Big[
(1+4x^2)\arctan2x-(1+8x^2+8x^4) \arctan x\Big] \bigg\}\,. \end{equation} 

\medskip

\noindent v) Hartree-diagram in Fig.\,3 with three medium insertions:

\begin{equation} \bar E_{n,4}(k_n) = {3g_A^4 M m_\pi^4 \over (4\pi f_\pi)^4 u^3}
\int_0^u \!dx x^2\int_{-1}^1 \!dy \Big[2uxy+(u^2-x^2y^2)H\Big]\bigg\{
{2s^2+s^4\over 2(1+s^2)}-\ln(1+s^2)\bigg\}\,. \end{equation}

\medskip

\noindent vi) Fock-diagram in Fig.\,3 with three medium insertions:

\begin{equation} \bar E_{n,5}(k_n) = {3g_A^4 M m_\pi^4 \over (4\pi f_\pi)^4u^3}
\int_0^u \!dx \bigg\{-{G^2\over 8}+{x^2\over 4}\int_{-1}^1\!dy \int_{-1}^1
\!dz {yz \,\theta(y^2+z^2-1) \over |yz|\sqrt{y^2+z^2-1}}\Big[s^2-\ln(1+s^2)
\Big] \Big[t^2- \ln(1+t^2)\Big]\bigg\}\,. \end{equation}

\medskip

\noindent vii)  Irreducible $2\pi$-exchange Hartree- and Fock-diagrams in
Fig.\,4: 
\begin{eqnarray} \bar E_{n,6}(k_n) &=& {m_\pi^5 \over(4 \pi f_\pi)^4}\bigg\{ 
\bigg[{1\over32u^3}( 83g_A^4+6g_A^2 -1) +{1\over 4u} (47g_A^4+2g_A^2-1) \bigg]
\,\ln^2(u+\sqrt{1+u^2}) \nonumber \\ && +\bigg[{u^2\over 30} (3g_A^4-50g_A^2-9)
-{u^4\over 15} (13g_A^4 +10 g_A^2+1) +{1\over 120} (1691g_A^4-90g_A^2 -73) 
\nonumber \\ && +{1\over16u^2}(1-6g_A^2-83g_A^4)\bigg] \sqrt{1+u^2} \ln(u+\sqrt
{1+u^2}) +{1\over32u} (83g_A^4+6g_A^2-1)\nonumber \\ && +{u\over 480} (397+210
g_A^2-11159g_A^4) +{u^5\over 1800}(119+710g_A^2+107g_A^4)  \\ && 
+{u^3\over 15}(9+55g_A^2-108g_A^4)+\bigg[ u^3\Big({1\over 3}+2g_A^2-5g_A^4
\Big)-{u^5\over 15} (13g_A^4+10g_A^2+1) \bigg] \ln{m_\pi\over 2\Lambda} \bigg\}
\nonumber \,.\end{eqnarray}
This expression includes the contributions from Hartree diagrams (not shown in
Fig.\,4) via the isovector central NN-amplitude $W_C(0)$ (see section 4.2 in
ref.\cite{nnpap1}). In terms of this amplitude the contribution of the
irreducible $2\pi$-exchange Hartree-diagrams is, $\bar E_{n}(k_n) = -\rho_n\,
W_C(0)/2$.

\medskip

\noindent viii) Power divergences specific for cut-off regularization:
\begin{equation} \bar E_{n,\Lambda}(k_n) = -{\Lambda \,k_n^3 \over 3(4\pi 
f_\pi)^4} \Big[2g_A^4 M +(3g_A^2+1)(g_A^2-1) \Lambda \Big] \,. \end{equation}
Note that compared to eq.(15) the attraction in neutron matter gets strongly
reduced. This welcome feature has its origin in the isospin dependence of the
$2\pi$-exchange.  

In the chiral limit, $m_\pi=0$, and truncated at order ${\cal
O}(k_f^4)$ the equation of state of neutron matter $\bar E_n(k_n)$ turns also
into the form eq.(1) with the coefficient $\beta_n = (g_{\pi N}/4\pi)^4 \,(27-
4\pi^2-24\ln2)/70 -(3/56)=-0.562$. The other coefficient $\alpha_n$ can be
easily read off from eqs.(33,39).  

\bigskip

\bigskip

\bild{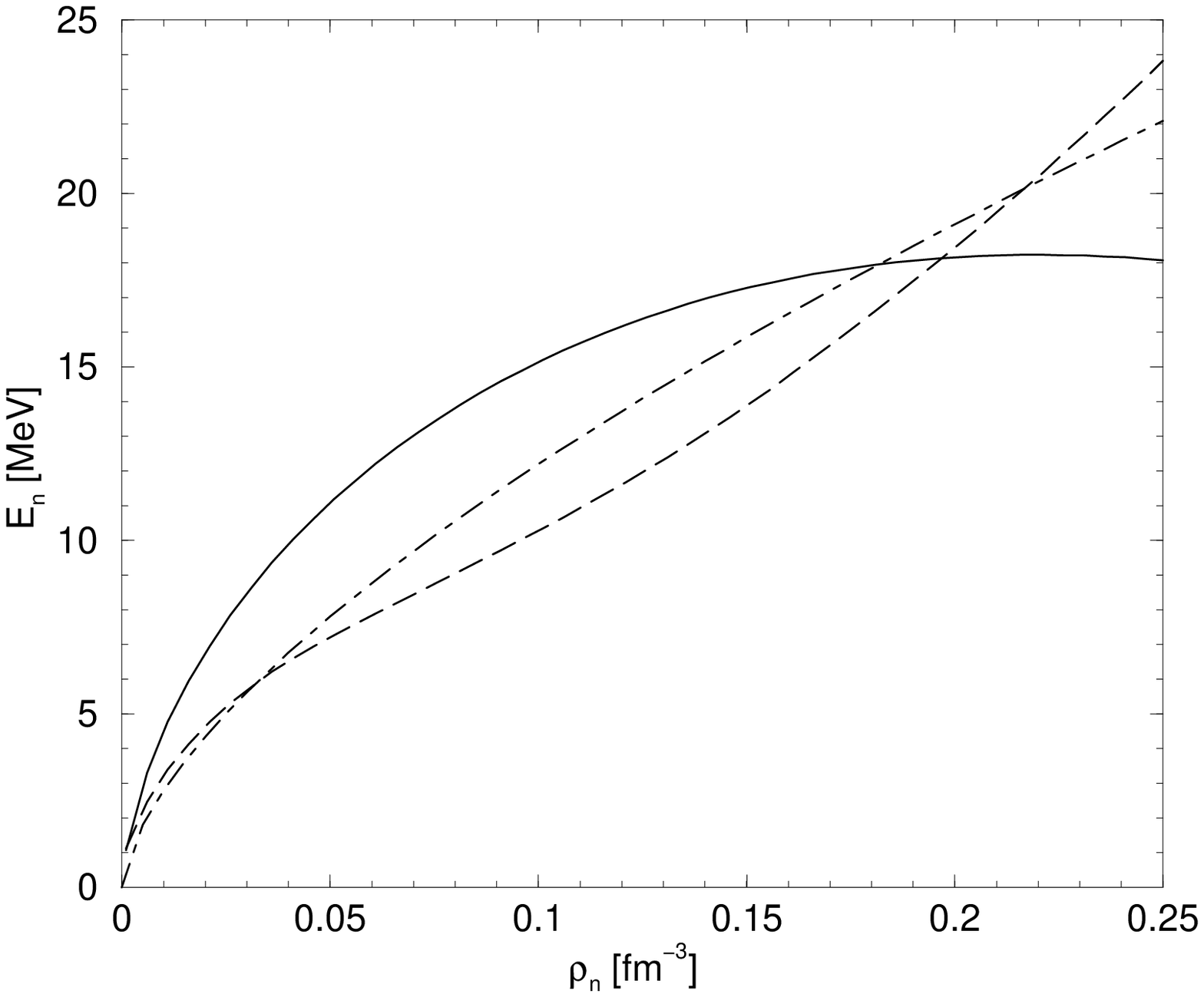}{10}
{\it Fig.\,8: The equation of state of pure neutron matter $\bar E_n(k_n)$. The
dashed line corresponds to the many-body calculation of ref.\cite{urbana}. The
dashed-dotted line represents eq.(40) with a neutron-neutron scattering
length of $a_{nn}=18.7\,$fm. The full line gives the result of chiral 
one- and two-pion exchange with a cutoff of $\Lambda =646\,$MeV.} 

\bigskip

If we use the same parameter input as in section 2.5, the resulting equation of
state of neutron matter, $\bar E_n(k_n)$, comes out as the solid line in
Fig.\,8. The convex shape of the curve is generic and does not change much with
the cut-off $\Lambda$. It demonstrates the limitations of our present
perturbative approach. The dashed line in Fig.\,8 corresponds to the many-body 
calculation of the Urbana group \cite{urbana}. This curve should be considered
as a representative of the host of existing neutron matter calculations
\cite{wiringa,glend,lirolf} which scatter around it. One observes a rough 
agreement up to neutron densities of about $\rho_n=0.25\,$fm$^{-3}$. At higher
densities our neutron equation of state starts to become unrealistic
because of its downward bending ($\beta_n<0$). This feature is inherited from
the asymmetry energy $A(k_f)$ (see Fig.\,7). Note however that our prediction 
for the neutron matter equation of state is still much better than 
the one of the $\sigma\omega$-mean field model \cite{chin} (for low neutron 
densities $\rho_n<0.25\,$fm$^{-3}$).  

The complete resummation of in-medium
multi-loop diagrams for a system with an unnaturally large scattering length
has been achieved in the limit of large space-time dimensions $D$ in 
ref.\cite{steele}. Neglecting the $1/D$-corrections (approximately less than 
$20\%$) the result of Steele \cite{steele} applied to neutron matter reads,    
\begin{equation} \bar E_n(k_n) = {k_n^2 \over M} \bigg[ {3\over 10 } - {a_{nn}
k_n \over 3\pi +6 a_{nn} k_n} \bigg] \,, \end{equation} 
with $a_{nn}$ the neutron-neutron scattering length. A recent measurement 
\cite{slaus} has found the precise value $a_{nn}=  (18.7 \pm 0.6 )\,$fm. (We 
use that sign-convention in which a positive scattering length corresponds to 
attraction.) The density dependence of eq.(40) is shown by the dashed-dotted 
line in Fig.\,8. The agreement with the many-body calculation of 
ref.\cite{urbana} is somewhat better.\footnote{We have assumed here that 
eq.(40) can be extrapolated to Fermi momenta as large as $k_n= 385\,$MeV.} 
Interestingly, the dashed-dotted line does almost not change if the
$nn$-scattering length is sent to infinity, $a_{nn}=\infty$, where $\bar
E_n(k_n) = 2k_n^2/15M$. 

Concerning  our neutron matter equation of state (solid line in Fig.\,8) one
should not forget that it has no free parameters (assuming that the high
momentum scale $\Lambda$ is the same as in isospin symmetric nuclear matter).
The fair agreement with the other two curves (for $\rho_n\leq 0.25\,$fm$^{-3}$)
is therefore quite surprising. The mere fact the neutron matter is predicted 
to be unbound is already non-trivial.

\section{Summary and Outlook}
The present work can be summarized as follows: We have used chiral 
perturbation theory to calculate the nuclear matter equation of state
systematically up to three-loops. The contributions to the energy per particle
$\bar E(k_f)$ from one- and two-pion exchange diagrams are ordered in powers of
the Fermi momentum $k_f$ (modulo functions of $k_f/m_\pi$).  We have evaluated
all contributions up-to-and-including order ${\cal O}(k_f^5)$. 

A momentum cut-off scale $\Lambda$ has been used to regularize the few 
divergent parts associated with chiral $2\pi$-exchange. The terms linear and
quadratic in $\Lambda$ effectively parametrize an equivalent zero-range 
NN-contact interaction which is strongly attractive in isospin symmetric
nuclear matter.  Its isospin dependence is revealed (and tested) in the 
asymmetry energy $A(k_f)$.  

The saturation mechanism behind the chiral $2\pi$-exchange is easy to 
understand (in the chiral limit $m_\pi=0$). An attractive $\alpha\, k_f^3/M^2$ 
term plus a repulsive $\beta\, k_f^4/M^3$ term, with proper coefficients 
$\alpha$ and $\beta$, lead automatically to a realistic nuclear
matter equation of state. 

Without inclusion of any further short-range terms, and adjusting the cut-off 
scale to the physically sensible value $\Lambda\simeq 0.65\,$GeV, the empirical
saturation point $(\rho_0 \simeq 0.17\,$fm$^{-3}, \bar E_0 \simeq -15\,$MeV) 
and the nuclear compressibility  $K\simeq 250\,$MeV are well reproduced.
Decomposing the binding energy $\bar E_0$ into chiral orders one recovers the
pattern of the realistic $(\alpha,\beta)$-parametrization. We note in passing
that the present calculation gives a single particle potential for a nucleon at
rest in equilibrium nuclear matter which amount to $U(0,k_{f0})\simeq
-53\,$MeV \cite{optical}. 

In the same framework and using the same parameters the density dependent 
asymmetry energy $A(k_f)$ has been calculated. The prediction for the 
asymmetry energy at the saturation point $A_0\simeq 34\,$MeV is in very good 
agreement with the empirical value. This is achieved again with just one 
cut-off $\Lambda$ instead of introducing additional contact terms: apparently,
one single cut-off scale is enough to represent the relevant short-distance 
dynamics. The downward bending of $A(k_f)$ at $\rho >0.2\,$fm$^{-3}$ presumably
indicates the limits of validity of the present chiral perturbation theory
calculation of nuclear matter.   

The equation of state of pure neutron matter as predicted in this framework is 
also in rough agreement with sophisticated many-body calculations and a
resummation result of effective field theory for neutron densities $\rho_n$ 
less than $0.25\,$fm$^{-3}$. The mere fact that neutron matter comes out  
unbound is non-trivial.

The present approach to the nuclear matter problem is quite different from most
other commonly used ones. We do neither start from a (so-called) realistic 
NN-potential which fits the deuteron properties and the NN-phase shifts, nor do
we employ relativistic mean field methods. Nevertheless, we find realistic 
nuclear binding, including its isospin dependence, already at the level of the
three-loop approximation of in-medium chiral perturbation theory. 

Questions remain, of course, whether the good results obtained so far
survive in higher orders. Relativistic $1/M$-corrections seem to present no 
problem since they are in general small (as long as the densities are not too 
high). Two-pion exchange with virtual $\Delta(1232)$-resonance excitation 
produces essentially all the medium-range scalar-isoscalar attraction which is
needed for the description of the peripheral NN-scattering \cite{nnpap2}. 
In the nuclear matter calculation the short-range contributions from
$2\pi$-exchange with $\Delta(1232)$-excitation as well as additional 
Pauli-blocking effects (related to diagrams with three medium insertions) come 
also into play. At the present stage, such effects are possibly hidden in the
cut-off scale $\Lambda$; at a higher level of "resolution", they can and have 
to be studied explicitly in the framework of chiral effective field
theory. Work along this line is in progress.   
\vspace{-0.3cm}
\subsubsection*{Acknowledgement}
We thank P. Ring for useful discussions.

\end{document}